\documentclass[preprint,12pt]{elsarticle}

\usepackage{epsfig}
\usepackage[utf8]{inputenc}
\usepackage[dvipsnames]{xcolor}
\usepackage[normalem]{ulem}
\usepackage{caption}
\usepackage{amssymb}
\usepackage{mathtools}
\usepackage{bbold}
\usepackage{amssymb,latexsym}
\usepackage{amsmath,amsbsy,bbm}

\newcommand{\eg}{\textit{e.g.}}
\newcommand{\ie}{\textit{i.e.}}

\newcommand{\ve}[1]{\ensuremath{\boldsymbol{#1}}}

\newcommand{\figref}[1]{fig.~\ref{#1}}

\definecolor{green}{HTML}{2E8B57}
\journal{PhysicsLetters A}

\begin{document}
\begin{frontmatter}

\title{Emergent four-body parameter \\
in universal two-species bosonic systems}

\author{Lorenzo Contessi}
\address{
  IRFU, CEA, Université Paris-Saclay, 91191 Gif-sur-Yvette, France
}

\author{Johannes Kirscher}
\address{
  Theoretical Physics Division, School of Physics and Astronomy,\\
  The University of Manchester, Manchester, M13 9PL, UK}
  
  \author{Manuel Pavon Valderrama}
\address{School of Physics,
Beihang University, Beijing 100191, China} 

\date{\today}

\begin{abstract} 
  The description of unitary few-boson systems is conceptually simple:
  only one parameter -- the three-body binding energy -- is required
  to predict
  the binding energies of clusters
  with an arbitrary number of bosons.
  Whether this correlation between the three- and many-boson
  systems still holds for two species of bosons for which only
  the inter-species interaction is resonant
  depends on how many particles of each species are in the system.
  For few-body clusters with species $A$ and $B$ and a resonant $AB$
  interaction, it is known that the emergent $AAB$ and $ABB$ three-body
  scales are correlated to the ground-state binding energies of the
  $AAAB$ and $ABBB$ systems, respectively.
  We find that this link between three and four bodies is broken
  for the $AABB$ tetramer whose
  binding energy is neither constrained by
  the $AAB$ nor by the $ABB$ trimer.
  From this de-correlation, we predict
  the existence of a scale unique
  to the $AABB$ tetramer.
  In our explanation of this phenomenon, we understand the $AABB$ and
  $AAAB$/$ABBB$ tetramers as representatives of two different universal
    classes of $N$-body systems with distinct renormalization-group and discrete-scaling properties.
\end{abstract}
\end{frontmatter}

\section{Introduction}
When a two-boson system is resonant, \ie, its scattering length is
considerably larger than the interaction range,
its behavior becomes independent of interaction details;
in this sense, it is {\it universal}.
Numerous instances of such systems are found in atomic, nuclear, and particle
physics~\cite{Braaten:2004rn}.
Resonant three-boson systems are even more intriguing as they display
a characteristic geometric bound-state spectrum.
This sequence of states was originally predicted by Efimov in the
seventies~\cite{Efimov:1970zz} and observed experimentally decades later
with ${}^{133}{\rm Cs}$ trimers~\cite{Kraemer:2006}.
The geometric factor between successive excited states in this spectrum
is universal and equal to $(22.7)^2$, while the ground-state binding energy ($B_3$) remains
sensitive to short-distance details of
the interaction~\cite{Bedaque:1998kg,Bedaque:1999ve} and
becomes a parameter of the theory.
If $B_3$ is known, the binding energies of larger
clusters can be predicted.
For the four-, five-, and six-boson systems at
unitarity, the relations $B_4 \sim 4.7\, B_3$, $B_5 \sim 10.1\, B_3$, and
$B_6 \sim 16.3\, B_3$, respectively, were found
numerically~\cite{Platter:2004he,Platter:2004zs,vonStecher:2009qw,vonStecher:2011zz}.

The occurrence of the Efimov spectrum is not exclusive to systems
composed of identical bosons.
Three-body systems with two species $A$ and $B$, \ie~ $AAB$ and $ABB$ 
(also referred to as {\it Tango} configurations~\cite{doi:10.1063/1.2710568}),
in which the two identical particles are bosons, do have a geometric spectrum, too,
even if only the $AB$ interaction is resonant.
This was experimentally observed for the first time with ${}^{87}{\rm Rb}$
and ${}^{41}{\rm K}$ trimers~\cite{PhysRevLett.103.043201}.
If the masses of species $A$ and $B$ are equal, the geometric factor is $(1986.1)^2$
(see \eg~\cite{Naidon:2016dpf}).
If the odd particle is lighter (heavier)
than the other two this factor is reduced (increased). 
For ${}^6{\rm Li}$ and ${}^{133}{\rm Cs}$, for example,
the theoretical prediction of a factor of $(4.9)^2$ for the ${\rm Li}\,{\rm Cs}\,{\rm Cs}$
trimer spectrum has been confirmed experimentally~\cite{PhysRevLett.112.250404}.

At this point, it is natural to wonder what happens with 
larger clusters composed of two species $A$ and $B$.
From the experience with
bosonic systems, the naive expectation is
that the ground-state energy of the two-species few-body clusters
is proportional to that of the trimer.
And indeed, this is the case for 
$AAAB$ clusters~\cite{PhysRevLett.108.073201,PhysRevLett.113.213201} %
(or $ABBB$ clusters, depending on the statistics of species $A$ and $B$).
 
In this manuscript, we investigate the $AABB$ system,
and find that its ground-state binding energy
cannot be predicted solely from the $AAB$ or $ABB$ three-body
parameters and a unitary $AB$ system.
This behavior was already foreshadowed in
the independent $AABB$ and $AAB/ABB$ limit cycles
in the Born-Oppenheimer $m(A)\ll m(B)$ limit~\cite{de_Paula_2020},
and our study establishes this decorrelation between three and
four-body systems for equal masses.

Furthermore, by treating the $ABBB$ and $AABB$ systems as representatives
of generic classes of $N$-particle systems in which only a certain
subset of pairs interacts resonantly with a contact $S$-wave potential
(all the other two-body pairs are non-interacting), we gain a more general understanding
of correlations amongst multi-species few-body clusters.
Our description has the advantage of arranging the different two-species
systems in two defined categories (which we call \textit{circle} and
\textit{dandelion}) independently of their physical realizations.
Based on numerical results, we conjecture that there is only
one finite scale associated with each of these categories.

\vspace{2mm}
\section{Theoretical framework:}

\begin{figure}
\begin{tabular}{cc}
 3-circle &  3-dandelion \\
 \includegraphics[width=0.40 \textwidth, trim=0 100 0 100, clip ]{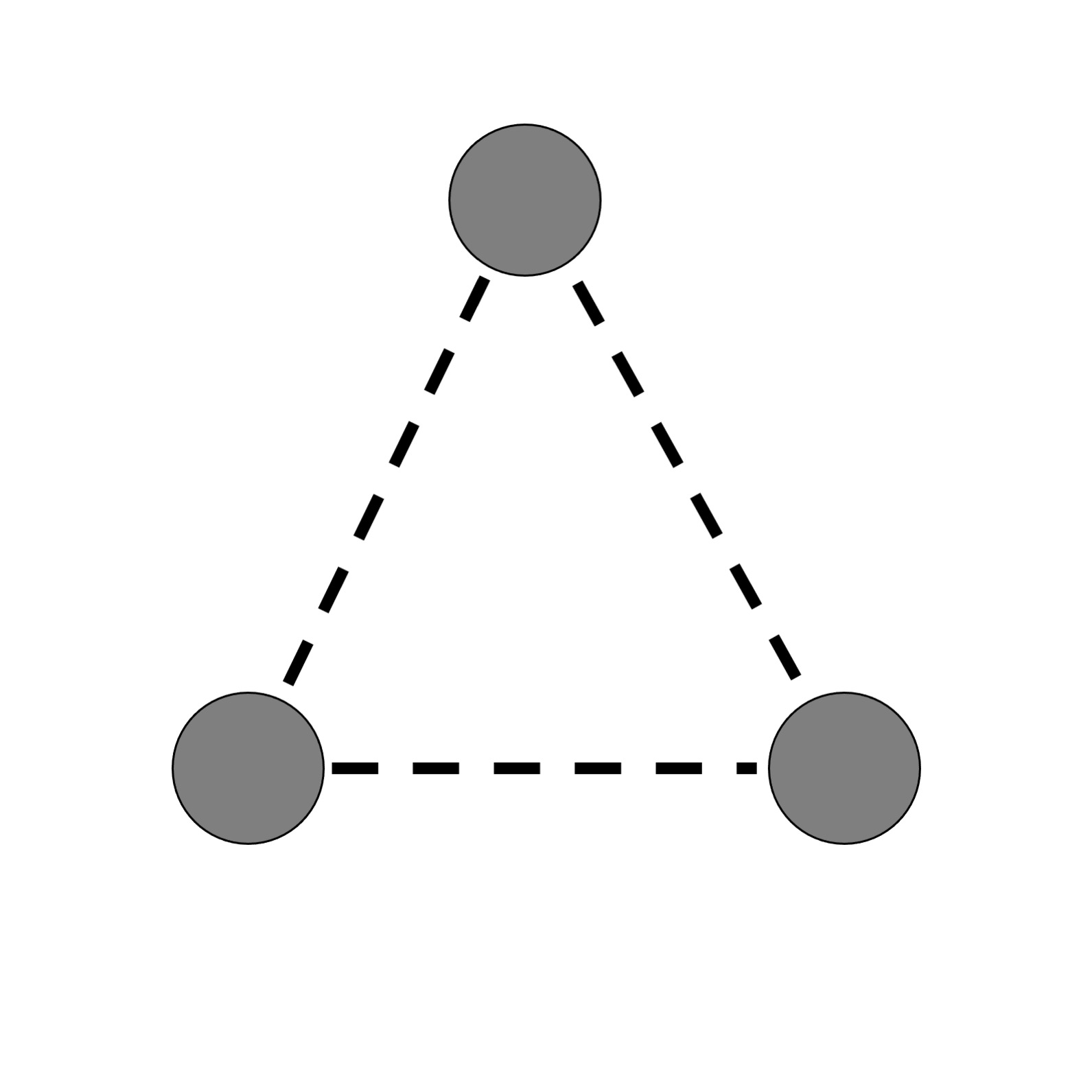} &   \includegraphics[width=0.40 \textwidth, trim=0 100 0 100, clip ]{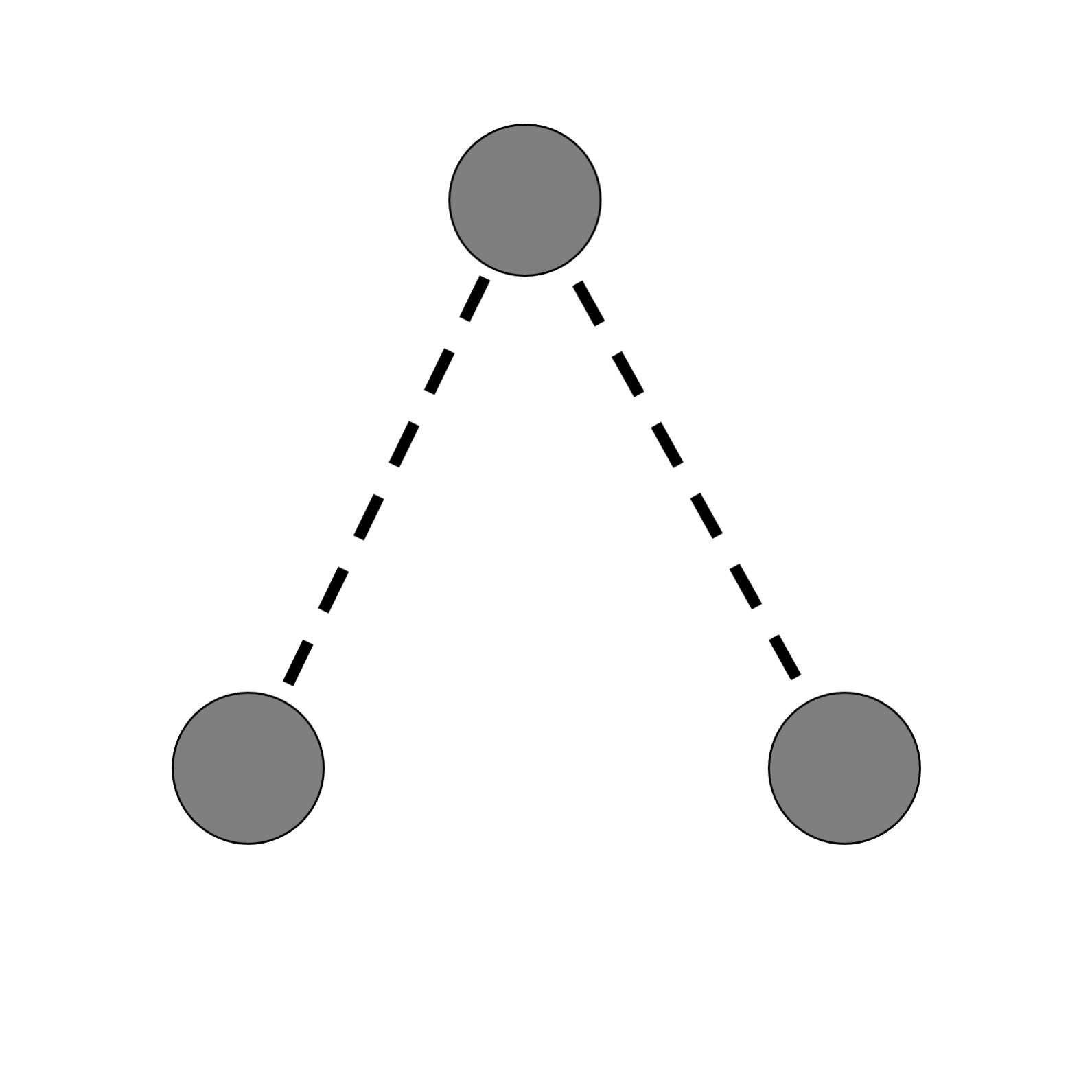} \\
 4-circle&  4-dandelion \\
 \includegraphics[width=0.40 \textwidth, trim=0 100 0 100, clip ]{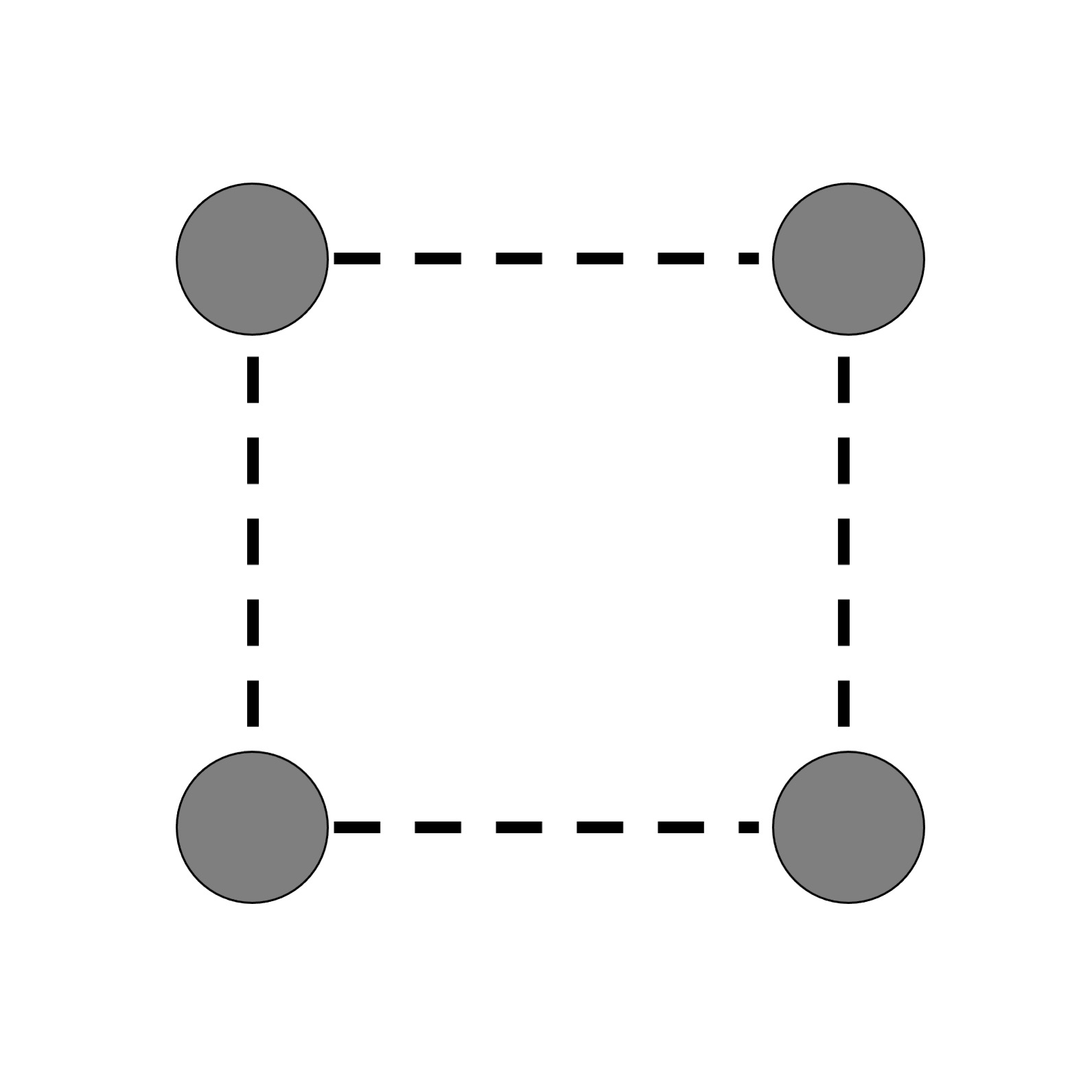} &   \includegraphics[width=0.40 \textwidth, trim=0 100 0 100, clip ]{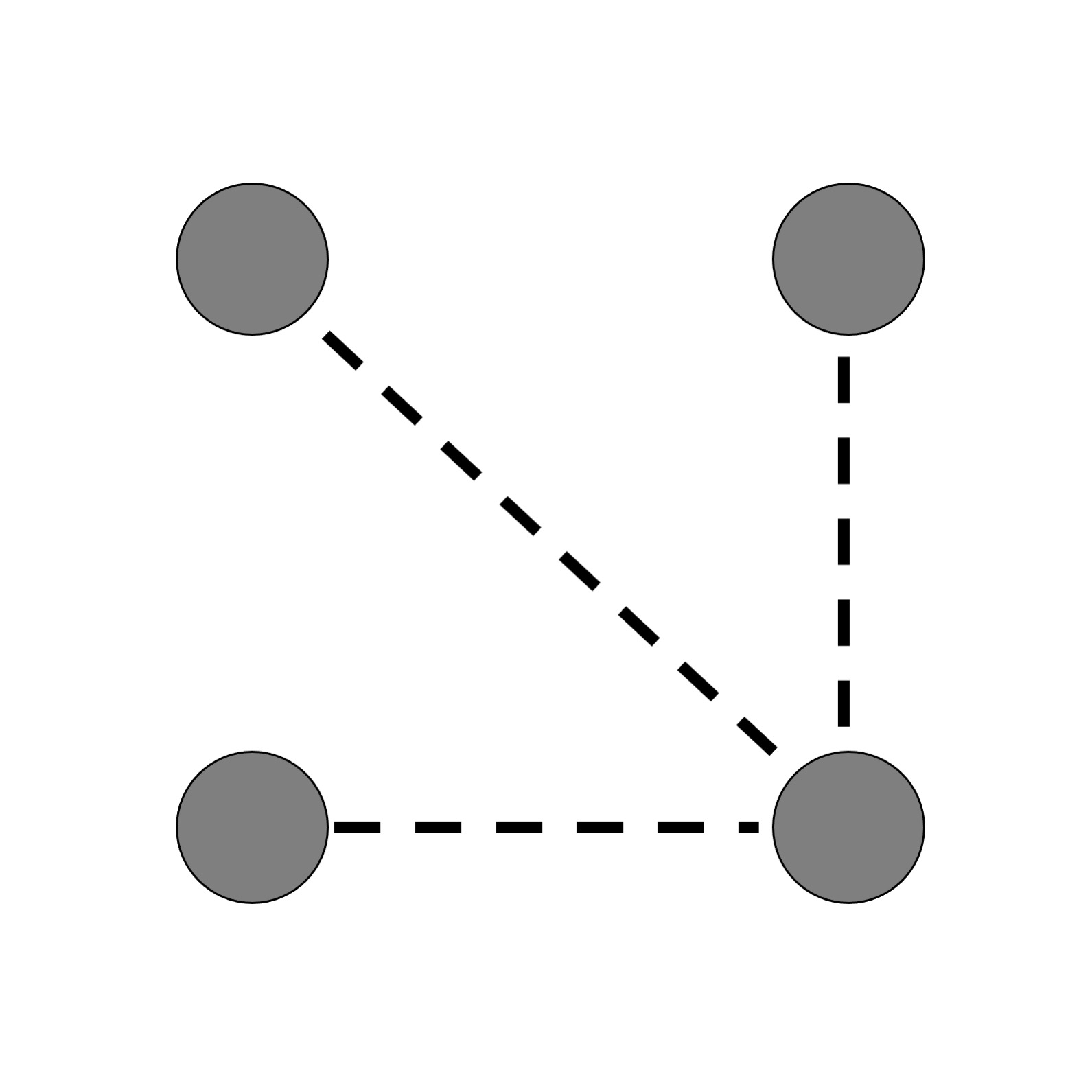} \\
 5-circle & 5-dandelion  \\
 \includegraphics[width=0.40 \textwidth, trim=0 100 0 100, clip ]{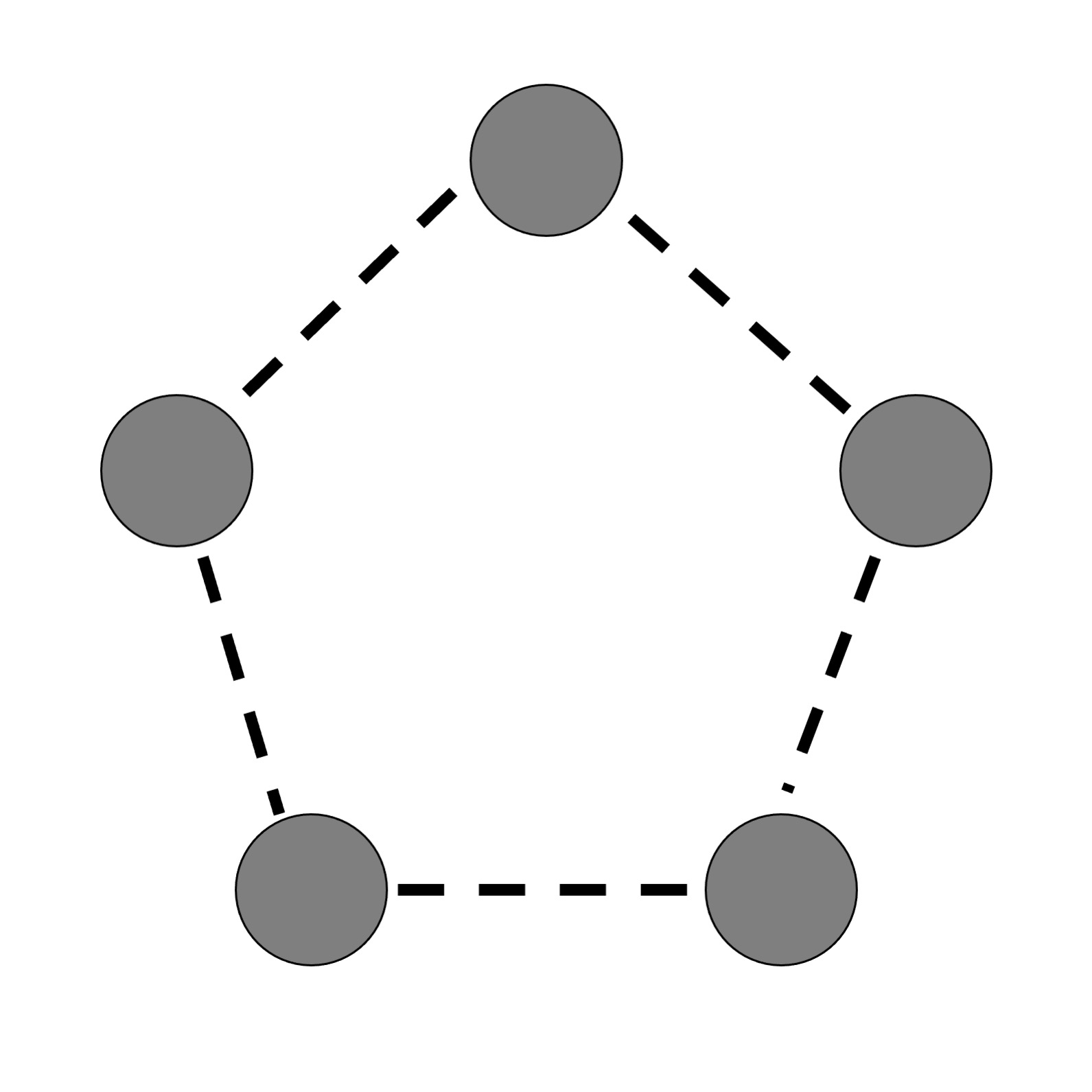} &   \includegraphics[width=0.40 \textwidth, trim=0 100 0 100, clip ]{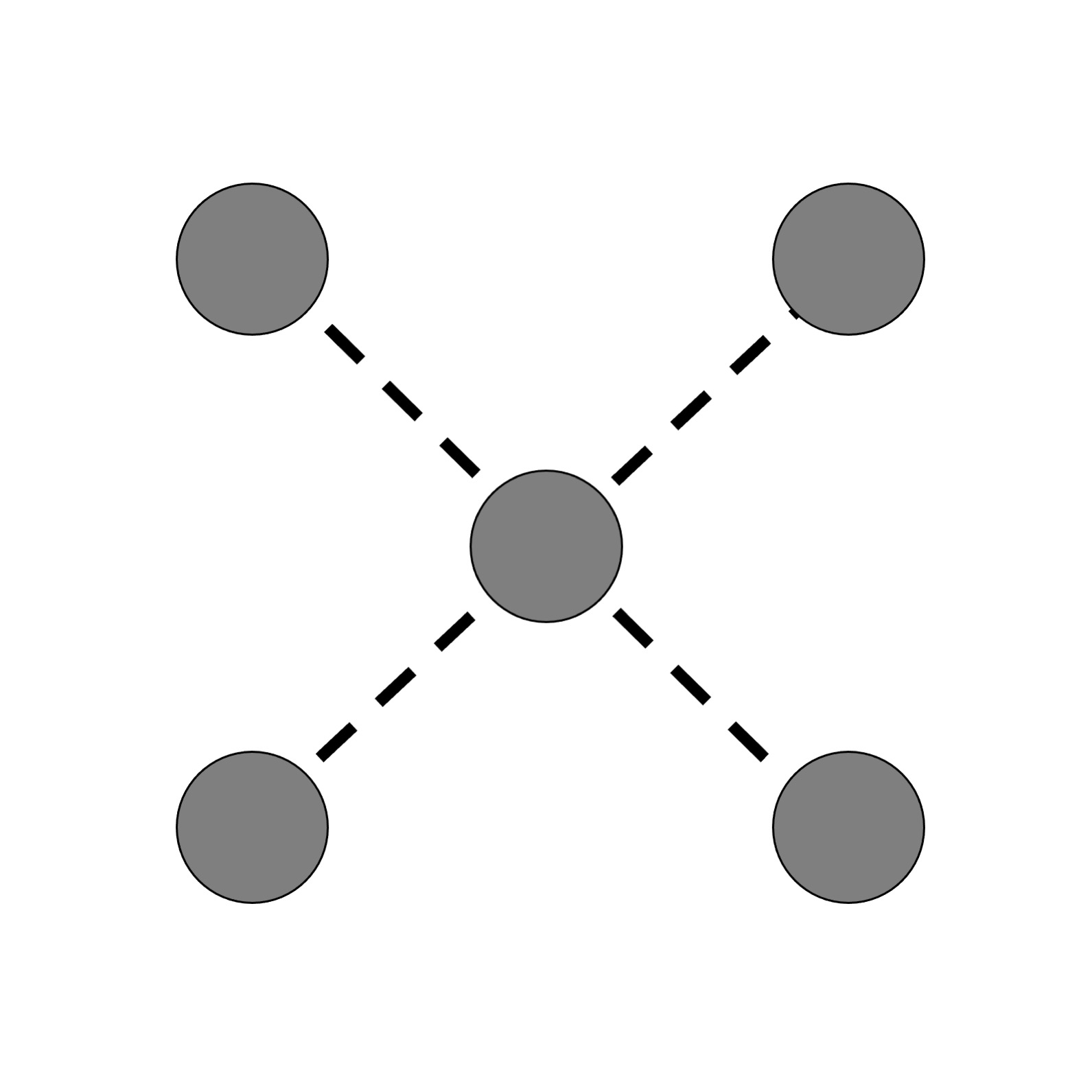} \\
\end{tabular}
\caption{Interaction graphs of two classes of $N$-body systems: the $N$-circle
  and the \mbox{$N$-dandelion.}
  Gray nodes represent particles and dashed lines/edges
  interactions between the connected particles.
}
\label{fig:systems}
\end{figure}

We will first give a definition of the categories and later
  relate them to the specific two-species bosonic systems.
We introduce the two classes (\figref{fig:systems}) for
distinguishable particles (the nodes) in which only
a specific subset of interactions is resonant (the dashed lines).
In the first class -- the $N$-circle --
each of the $N$ nodes/particles interacts resonantly
with exactly two neighbors.
In the second class -- the $N$-dandelion -- a central particle
interacts with all remaining $(N-1)$ particles that do not interact
amongst each other.

The quantum-mechanical nonrelativistic dynamics of these graphs
obey the $N$-body Schr\"odinger equation
\begin{equation}\label{eq.FBprob}
  \left({H}_0+{V}\right)\vert 1,2,\ldots,N\rangle =
  E_N\,\vert 1,2,\ldots,N\rangle\;\;,
\end{equation}
with an equal-mass kinetic term ${H}_0$ acting on a state with energy $E_N$ 
(for bound state solutions, $- B_N:= E_N < 0 $).
The shape of the graph is encoded in the potential ${V}$, which specifies
the subset of all the $N(N-1)/2$ interaction pairs $(i,j)$
that are resonant.
In particular, the potentials defining the \mbox{$N$-circle} and $N$-dandelion are
\begin{eqnarray}
&{V}^{N\textrm{-circle}}&=\sum_{i=1}^{N}\,{V}(i,(i\,\textrm{mod}\,N)+1)\, ,\\
&{V}^{\textrm{N-dandelion}}&=\sum_{i=1}^{N-1}\,{V}(i,N) \; \; .   
\end{eqnarray}
For the resonant interaction between equal-mass particles, we use a
contact potential regularized with a Gaussian cut-off function:
\begin{eqnarray}
  V_{2}(\ve{r}; R_c) = C(R_c)\,\delta^{(3)}(\ve{r}; R_c) \, , \\[2mm]
  \delta^{(3)}(\ve{r}; R_c) = \frac{e^{-(r/R_c)^2}}{\pi^{3/2} R_c^3} \, ,
  \label{eq:delta-2B-regularization}
\end{eqnarray}
which is the leading-order of an effective field
theory for systems with large scattering length~\cite{vanKolck:1998bw}.
In order to prevent the three-body collapse
(\ie~ $B_3 \to \infty$ when ${R_c \rightarrow 0}$~\cite{Thomas:1935zz}),
we introduce a zero-range three-body potential~\cite{Bedaque:1998kg}
adopting the regulator prescription from the pair interaction:
\begin{eqnarray}
  V_{3}(\ve{r}_{123}; R_c) =
  D(R_c)\sum_{cyc}\,\delta^{(3)}(\ve{r}_{12}; R_c)\,\delta^{(3)}(\ve{r}_{23}; R_c) \, .
  \label{eq:delta-3B-regularization}
\end{eqnarray}

  Numerically, we realize the zero-range limit with $R_c \in [0.01,1]$
  for dandelions and $R_c \in [0.1,10]$ for circles.
  At the lower end of these ranges the binding energies show
  relatively little dependence on $R_c$.
  In order to approximate a zero-range interaction we choose the
  minimum value of the cut-off as to be much smaller than any
  other length scale in the systems we study (yet it cannot be reduced
  much further without impairing the numerical stability of
  the calculations).
The two-body coupling strength $C(R_c)$ was calibrated via
  the Numerov algorithm to approximate a unitary scattering
  length $|1/a_0| < 10^{-5}$. In addition to $\hbar=c=1$,
  we set $m=1$.
We chose a single three-body scale for the 3-circle and the 3-dandelion,
  namely the ground-state binding energy $B_3=0.01$
  \mbox{($1/a_0\ll B_3\ll m$)}.
  Accordingly, we renormalized the strength of $D(R_c)$
  in both systems to reproduce this value of $B_3$.
This results in two sets of three-body
  couplings $D(R_c)$ representing a repulsive potential,
  where we note that
  setting the 3-circle energy to $0.01$ demands a stronger repulsion
  than fixing the 3-dandelion's ground state to $0.01$.

\begin{figure}
\begin{tabular}{ccc}
 \includegraphics[width=0.29 \textwidth, trim=0 0 0 0, clip]{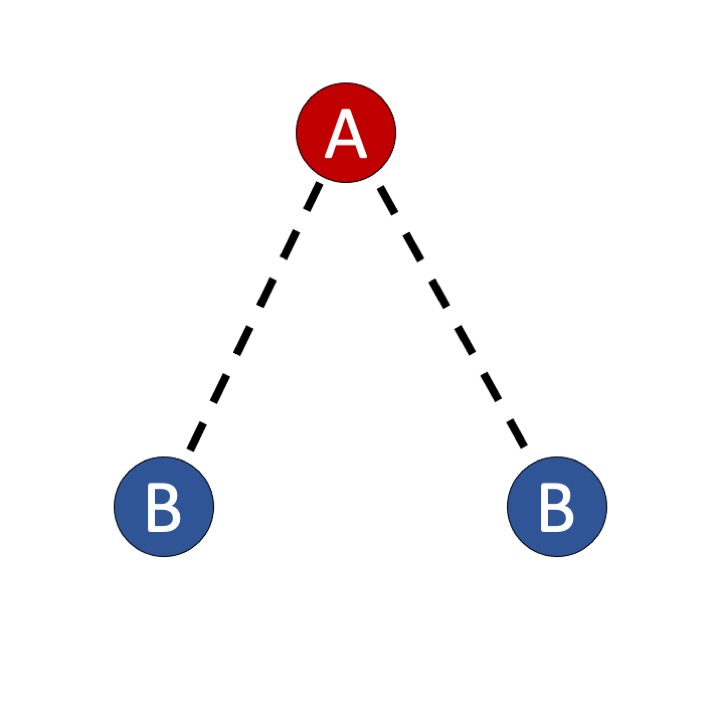} &   \includegraphics[width=0.29 \textwidth, trim=0 0 0 0, clip]{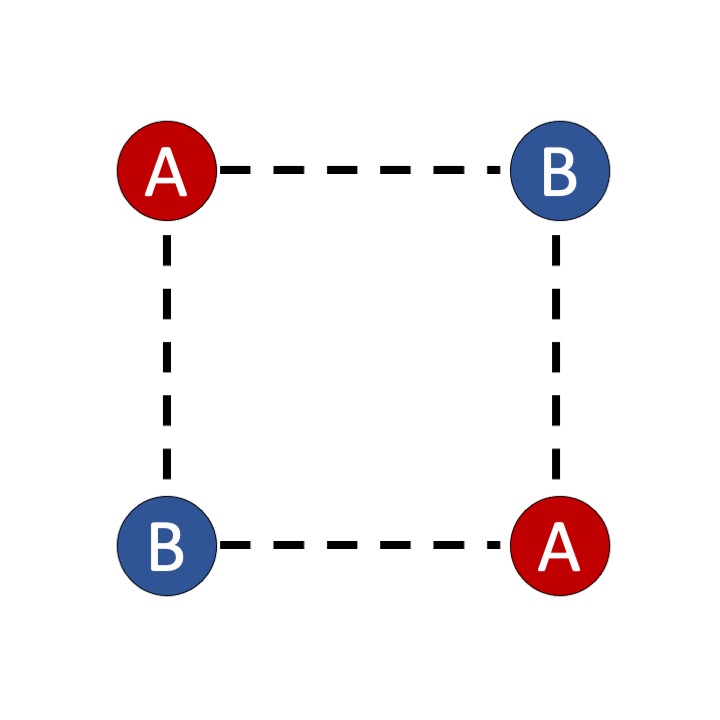} &
 \includegraphics[width=0.29 \textwidth, trim=0 0 0 0, clip]{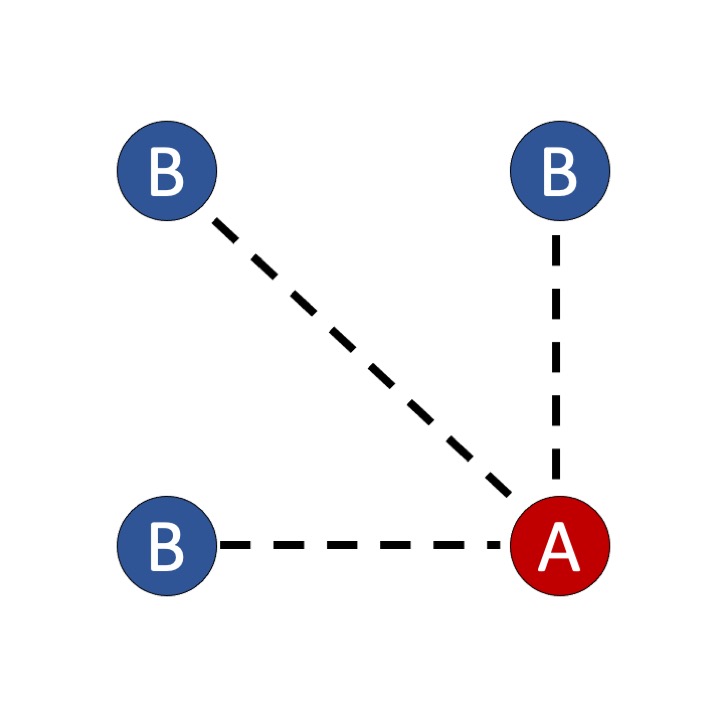}    
\end{tabular}
\caption{
Realizations of the circle and dandelion classes
  with two-species of bosons ($A$ and $B$).
}
\label{fig:systems_AB}
\end{figure}
The two interaction classes can be realized
  as two-species bosonic clusters
in which only the inter-species interactions are resonant.
Stated explicitly, the $ABB$ and $ABBB$ ground states
are $3$- and $4$-dandelions, respectively, while the $AABB$ cluster
is a $4$-circle (see \figref{fig:systems}~and \figref{fig:systems_AB}).
The $3$-circle is equivalent to the standard three-boson system 
at unitarity~\footnote{In our notation, the 3-circle would be
    equivalent to the $AAA$ system if same-species interactions
    were non-zero.
    If the latter were resonant too, the $AAA$ (\ie~3-circle) system
    would exhibit the Efimov effect~\cite{Efimov:1970zz}.
    If we follow the convention used in this work of setting
      same-species interactions to zero, the 3-circle will be
      equivalent to an $ABC$ system, with $B,C$ representing different
      particle species (while the $AAA$ system would be a free system).
      This system (\ie~three distinguishable particles
        in the unitary limit) shows the same type of geometric Efimov spectrum
        as the standard three boson system~\cite{Naidon:2016dpf}.
    %
    %
},
but it is no subsystem of the two-species clusters considered here.
We advance this is why $AABB$ exhibits its own characteristic
four-body scale.

This equivalence is explained by noting that
the ground state and the potential have the same symmetries, \ie,
cyclic permutations for the circle and permutations of
the $(N-1)$ non-interacting particles for the \mbox{$N$-dandelion.}
This implies that these particles effectively behave as indistinguishable,
though in principle not necessarily as bosons.
Yet, the contact interactions \eqref{eq:delta-2B-regularization}~and \eqref{eq:delta-3B-regularization} enforce
bosonic behavior for the $N=3,4$ circles and the $N=3,4,5$ dandelions.
In the $R_c \to 0$ limit, these interactions are only non-vanishing
in relative $S$-wave configurations between interacting pairs,
which requires them to behave as bosons.
If the non-interacting pairs are antisymmetric, this will force 
the interacting pairs to be antisymmetric too~\footnote{
This follows from permutation algebra: labeling the particles
according to species ($A$ or $B$ in \figref{fig:systems_AB}),
any permutation involving $AA$ or $BB$ pairs must be expressed as
the product of an odd number of permutations involving $AB$ pairs.
Sign consistency of permutations acting on the wave function
thus requires the sign of $AA$/$BB$ and $AB$ exchanges
to be the same.
},
resulting in a vanishing contribution from the contact-range interaction.
As a consequence, provided that the interaction is of a contact-range type,
all pairs behave symmetrically under pair permutations in the ground state.
This is the reason why
our calculations employ states with distinguishable particles.
This choice allows us to avoid the explicit (and unnecessary)
  symmetrization of the numerical wave function, thus reducing
  the number of components to be computed. This in turn makes
  the computation of larger cut-off-radii ensembles feasible.
However, we were careful to verify the
equivalence numerically for selected three- and four-body cases.\footnote{
Specifically, we benchmarked the 3- and 4-circle systems at $R_c=0.25$
and the 3- and 4-dandelions at $R_c=0.1$. Furthermore,
the 4-circle was tested for $0.17<R_c<2$.}

All three- and four-body results obtained for this work employ
the stochastic variational method \cite{Suzuki:1631377}
and were bench-marked for a sample set of three- and four-body predictions
with the refined-resonating-group method~\cite{Hofmann:1986}.
Within this numerical framework, we calculate ground-state energies of 
three-, four-, and five-body representatives of the circle and dandelion.
The convergence of these energies to a finite value indicates
the renormalizability of the theory for the respective systems.
The collapse of the ground states, \ie~ $B_N \to \infty$ for $R_c \to 0$,
is, in turn, the signature of an emergent new scale.

\vspace{2mm}
\section{Results:}
\begin{figure}[th!]
 \includegraphics[width=\textwidth]{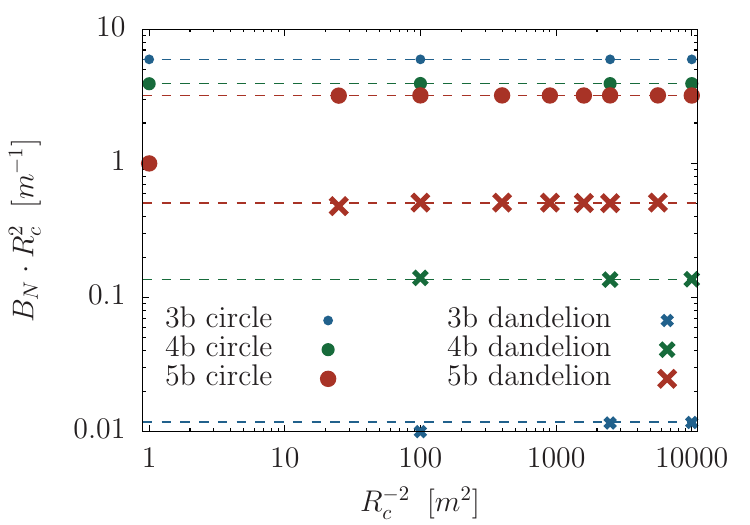}
 \caption{
 Cut-off dependence of ground-state binding energies ($B_N$) for
 trimer, tetramer, and pentamers with an interaction given by
 resonant two-body potentials which realize circular and dandelion
 systems. Results are obtained without a three-body counterterm.
In showing $B_N R_c^2$, the
divergent behavior of the binding energy is highlighted.
 }
\label{fig:circle_collapsing}
\end{figure}
\begin{figure}[th!]
\includegraphics[width=\textwidth]{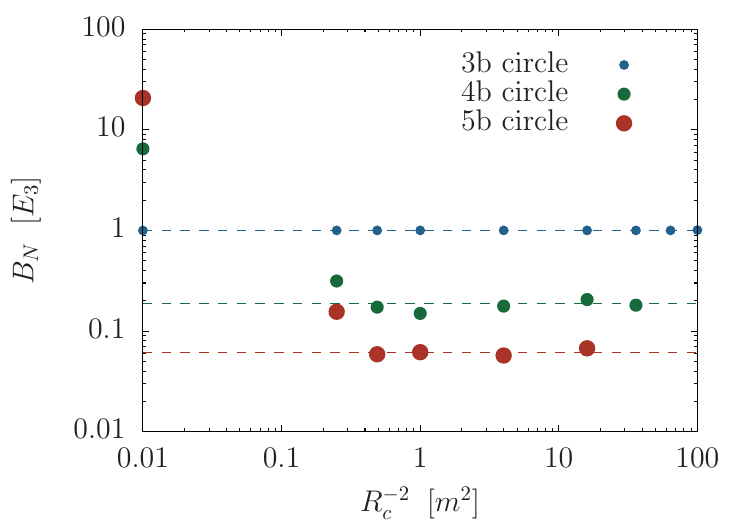}
\caption{Convergent behavior of the $3$-, $4$- and $5$-circle ground-state binding energies
with an appropriately renormalized three-body repulsion ({\it cf.}~\figref{fig:circle_collapsing}
for unrenormalized results). }
\label{fig:circle_converging}
\end{figure}
\begin{figure}[th!]
 \includegraphics[width=\textwidth]{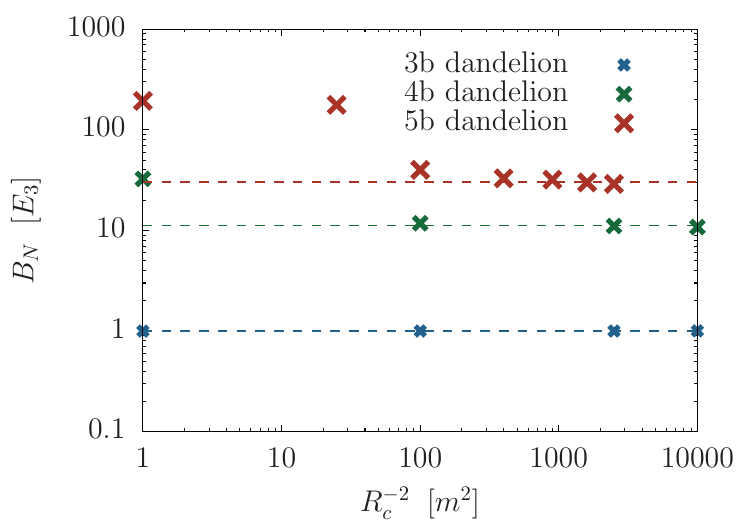}
\caption{Convergent behavior of the $3$-, $4$- and $5$-dandelion ground-state binding energies
with appropriately renormalized three-body repulsion ({\it cf.}~\figref{fig:circle_collapsing}
for unrenormalized results).}
\label{fig:dandelion_converging}
\end{figure}

We begin our analysis recovering the expected collapse associated with the original Efimov spectrum.
Specifically, we demonstrate that in the absence of a three-body repulsion all
considered systems collapse as $B_N \propto 1/R_c^2$
(see \figref{fig:circle_collapsing}).
Then, we analyze the renormalizability of the $N=4,5$ dandelion and circle
with a contact two-body resonant interaction and a contact three-body repulsion
that stabilizes the respective trimers.
In the large cut-off limit, we find the energy of these systems stable against
variations of the cut-off and approximately
\mbox{$B_4^{\rm dand} / B_3^{\rm dand} \simeq 11$} and
\mbox{$B_4^{\rm circ} / B_3^{\rm circ} \simeq 0.2$}
for the four-body dandelion and circle respectively, and 
\mbox{$B_5^{\rm dand} / B_3^{\rm dand} \simeq 30$} and
\mbox{$B_5^{\rm circ} / B_3^{\rm circ} \simeq 0.06$}
for the five-body dandelion and circle. 
The cut-off dependence of these ratios is shown
in \figref{fig:circle_converging}~and \figref{fig:dandelion_converging}.

The $4$- and $5$-dandelion binding energies are large
when compared to that of the $3$-dandelion, but the reason
why this is the case remains elusive.
For circular systems, the absolute binding energy decreases
with each additional particle on the circle.
Whether the binding energy of the circle converges to zero with the
  number of particles going to infinity remains an open question.
Nevertheless, the circles are particle-stable because no $(N+1)$-body
circle can decay into an $N$-body circle.
Although not obvious, we find the closest decay threshold to be
the complete disintegration of the systems.

\begin{figure}[th!]
 \includegraphics[width=\textwidth]{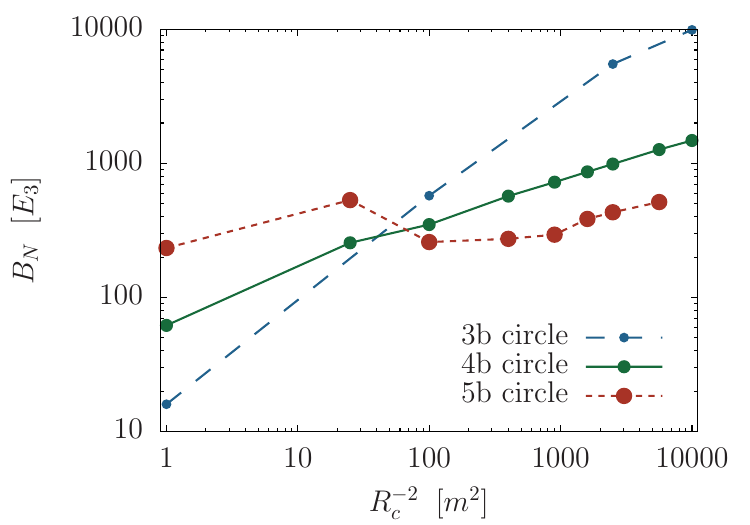}
 \caption{Divergent behavior of the $3$-, $4$- and $5$-circle ground-state binding energies
with a three-body repulsion inconsistently renormalized to a $3$-dandelion ({\it cf.}~\figref{fig:dandelion_converging}
for the results based on the properly running counter term).
 }
\label{fig:circle_less_collapsing}
\end{figure}

Next, we assess whether the three-body force that renormalizes the
$3$-dandelion suffices to renormalize the $4$- and $5$-circle too.
The binding energy of the dandelion-regularized circle
shows no sign of convergence to a particular value.
However, its increase with the cut-off is less steep relative
to the case without three-body repulsion
(see \figref{fig:circle_less_collapsing}).
The collapse was tested up to a cut-off of $100$ (in units of mass), and
although we cannot numerically exclude a stabilization at
some even larger cut-off, we deem such behavior unlikely for two reasons:
first, the extremely large binding of these systems for the considered range
of cut-offs suggests that any hypothetical finite limit has a highly unnatural
energy which, {\it a priori}, is not justified by any physical reason.
Second, the fact that the circle is correctly
renormalized via the repulsion
set by the $3$-circle means that it cannot be simultaneously renormalized
by the repulsion that stabilizes the $3$-dandelion.
The discrete scaling factors 
of the $3$-dandelion and $3$-circle are {\it de facto} different,
being respectively
$1986.1$ and $22.7$
for equal mass systems~\cite{Naidon:2016dpf}.
These results indicate that, in the unitary limit, the dandelion and circle
categories represent two different types of universal behavior.

It is helpful to consider the consequences for two-species
three-boson systems (either $AAB$ or $ABB$) to which one additional
boson is added. For an inter-species two-body potential at resonance,
the $AAAB$/$ABBB$ mixtures -- the dandelions -- exhibit bound states
which are tied to those of the $AAB$/$ABB$ three-boson spectrum.
The $AABB$ system, in contrast, is not renormalized by the same three-body
repulsion.
It obtains its renormalization condition from the $3$-circle, 
which cannot be built in terms of $A$/$B$-mixtures.
This three-body potential cannot be included in the theory a posteriori,
since it would spoil the $ABB$ system renormalization. 
Therefore, the renormalization of the $AABB$ tetramer requires the introduction of its own four-body force at leading order (in contrast with the four-boson system, in which it is subleading~\cite{PhysRevLett.107.135304,Bazak:2018qnu,Frederico:2019bnm}).

It is noteworthy that the $5$-circle cannot be represented by $A$/$B$-mixtures,
and we thus abstain from a more detailed analysis of this system.
  However, it belongs to the same universality class as the $4$-circle and
  is also renormalized by the $3$-circle three-body force
  (see \figref{fig:circle_converging}).
  From this, we would expect the $5$-circle to be renormalized by
    the four-body force that renormalizes the $4$-circle
    in its physical representation (\ie~the $AABB$ tetramer),
    though for the moment this is merely a conjecture.


The experimental verification of our theoretical results is
complicated by the fact that most physical $AABB$ systems
have mass imbalances.
The heteronuclear $^{85}\rm{Rb}$-$^{87}\rm{Rb}$ system (both bosons), for which
a Feshbach resonance has been observed~\cite{PhysRevLett.97.180404}, might
be an exception. It is relatively close to the equal-mass limit and
might enable (albeit experimentally challenging) a measurement of
$B_4^{AAAB} / B_3^{AAB}$ and $B_4^{AABB} / B_3^{AAB}$ ratios.
To make a comparison we would need a second system with similar features, but no such system is known experimentally. Hence the identification of an independent four-body scale remains elusive.
Yet, for multi-species systems with similar mass imbalances, such as
$^{41}\rm{K}$-$^{87}\rm{Rb}$~\cite{PhysRevLett.103.043201,Wacker_2016},
$^{87}\rm{Rb}$-$^{133}{\rm Cs}$~\cite{PhysRevA.89.033604}
and $^{23}{\rm Na}$-$^{39}{\rm K}$~\cite{PhysRevA.99.032711}
(for which the mass imbalances are $0.47$, $0.65$, and $0.59$, respectively),
the comparison of their $B_4^{AABB} / B_3^{AAB}$ ratios
might very well be meaningful.
If these ratios are wildly different, it could represent an experimental
verification of the existence of a four-body parameter in the $AABB$ system.
Another way to realize the equal-mass
dandelion and circular
systems is provided by Feshbach resonances between atomic hyperfine levels.
Our results
could then be confirmed if besides~${}^{87}$Rb~\cite{PhysRevA.82.033609}~a
second atomic species could be tuned to such a resonance and that the trimer
and tetramer energies within the respective condensates are experimentally
accessible.

\vspace{10mm}
\textbf{Acknowledgments:}
We thank  B.~Bazak, M.~Birse, T.~Frederico, and V.~Som\`{a}
for comments and discussions.
This work is partly supported by the National Natural Science Foundation
of China under Grants No.  11735003, No. 11975041. the Fundamental
Research Funds for the Central Universities and the Thousand
Talents Plan for Young Professionals.
M.P.V. thanks the IJCLab of Orsay for its long-term hospitality.
L. Contessi acknowledges the support and the framework of the "Espace de Structure et de réactions Nucléaires Théorique" (ESNT, http://esnt.cea.fr ) at CEA.

\bibliographystyle{elsarticle-harv}
\bibliography{broken.bib}

\end{document}